\documentclass[journal]{IEEEtran}

\usepackage{amsmath,amssymb,amsfonts}
\usepackage{algorithm}
\usepackage{algorithmic}
\usepackage{graphicx}
\usepackage{tikz}
\usetikzlibrary{arrows.meta,positioning,calc,fit,backgrounds,shapes.geometric}
\usepackage{textcomp}
\usepackage{xcolor}
\usepackage[colorlinks=true,allcolors=blue]{hyperref}

\newcommand{\bh}{\mathbf{h}}

\newcommand{\bR}{\mathbf{R}}
\newcommand{\bI}{\mathbf{I}}
\newcommand{\bW}{\mathbf{W}}
\newcommand{\bP}{\mathbf{P}}
\newcommand{\bF}{\mathbf{F}}
\newcommand{\by}{\mathbf{y}}
\newcommand{\bn}{\mathbf{n}}
\newcommand{\bw}{\mathbf{w}}
\newcommand{\bz}{\mathbf{z}}
\newcommand{\bB}{\mathbf{B}}
\newcommand{\bmu}{\boldsymbol{\mu}}
\newcommand{\bSig}{\boldsymbol{\Sigma}}
\newcommand{\CN}{\mathcal{CN}}

\newcommand{\bbC}{\mathbb{C}}

\DeclareMathOperator*{\argmin}{arg\,min}
\DeclareMathOperator*{\argmax}{arg\,max}

\begin{document}

\title{Active Inference for Joint Port Selection and Pilot\\
       Allocation in Fluid Antenna Systems Under Partial CSI}

\author{Kian~Fotovat$^{1}$, Kamran~Fotovat$^{2}$, and Zijun~Wang$^{3}$%
  \thanks{$^{1}$Dept.\ of Electrical and Computer Engineering, University of
  Tehran, Tehran, Iran.}%
  \thanks{$^{2}$Iran University of Science and Technology, Tehran, Iran.}%
  \thanks{$^{3}$Dept.\ of Electrical Engineering, University at Buffalo, The
  State University of New York, Buffalo, NY, USA.}%
}

\maketitle

\begin{abstract}
Fluid antenna systems (FAS) obtain spatial degrees of freedom by activating a
few radiating ports from a dense grid of candidates, but deciding which ports
to activate presupposes channel knowledge at ports that are never measured.
Prior port-selection work assumes full or near-full channel state information
(CSI) and treats the pilot overhead as a fixed cost. We instead formulate port
selection and pilot allocation \emph{jointly} as active inference. A
spatio-temporal generative model --- sub-wavelength spatial correlation
together with an order-$p$ autoregressive temporal prior --- maintains a
Gaussian belief over all $N$ port channels, and the transmitter chooses both
the ports it serves and the subset of them it pilots by minimizing an expected
free energy that trades achievable rate against information gain and
port-switching cost. Selection is greedy and submodular, at $\mathcal{O}(NM)$
per slot. On a $441$-port grid with a hybrid front end at
$n_{\mathrm{RF}}=2K$, the agent attains 91\% of a full-CSI genie's sum rate
while measuring only 2.3\% of the candidate ports; pilots can then be withdrawn
from ports that are still served, one fifth of them for 2\% of the sum rate and
two fifths for 8\%. No training is required.
\end{abstract}


\section{Introduction}
\label{sec:intro}

Fluid antenna systems (FAS), also called movable- or reconfigurable-position
antenna systems, replace fixed radiating elements with elements whose position
can be adjusted among a dense set of candidate positions, or ports, within a
compact aperture \cite{wong2021fas,fas_tutorial}. By moving to a favourable point of the
spatial fading pattern, an FAS harvests diversity that a fixed array of the
same size cannot, using only a handful of radio-frequency (RF) chains, since at
any instant only the activated ports are connected to the front end. The
pivotal control problem is port selection: deciding slot by slot which subset
of the candidates to activate.

A port can be scored only if its channel is known, so conventional selection
presupposes channel state information (CSI) at every candidate. That is
expensive. The aperture is sampled at sub-wavelength spacing, so the ports are
packed far more densely than a half-wavelength array and the number of
coefficients to estimate grows rapidly with aperture. Worse, only activated
ports can be observed at all. The bottleneck is therefore not the selection
arithmetic but channel acquisition under a sensing budget: the system must
decide which ports are worth measuring --- and, we add, whether a port that is
served needs to be measured at all.

Several lines of work approach this and leave the gap open. Learning-based
selection --- deep reinforcement learning over ports
\cite{drl_switching,fas_isac_drl} and online bandits \cite{online_portsel} ---
learns policies directly, but treats the full channel as an observed, cost-free
input and requires extensive training. Channel estimation for FAS
\cite{sbar,fas_lmmse,fas_oversampling} exploits spatial correlation to
reconstruct the channel from few pilots, and predictive selection extends this
across time, extrapolating the port channels forward from a fixed set of
observed ports with an LSTM or a Gaussian process
\cite{fas_fastport,fas_switchdelay}. Both solve estimation in isolation: they
recover or predict the channel without deciding what to sense or what to serve,
fix in advance which ports are measured, and never weigh a measurement against
its cost. What is missing is a
formulation in which sensing and serving are chosen jointly under an explicit
budget.

Active inference supplies that perspective. Rooted in the free-energy principle
\cite{parr2022active}, it casts a decision-maker as an agent holding a
generative model of its environment and acting to minimize an expected free
energy (EFE), which decomposes into a pragmatic value rewarding goal attainment
and an epistemic value rewarding informative actions
\cite{friston2015active}. The fit to FAS is close:
the hidden state is the channel at unmeasured ports, the generative model is
its spatial and temporal correlation, the pragmatic value is throughput, and
the epistemic value is what a pilot reveals about the ports left unseen. The
exploit--explore tension that defines the problem is thus expressed by the EFE
rather than engineered in by hand. Our contributions are as follows.

\begin{enumerate}
\item We formulate port selection \emph{and pilot allocation} jointly as active
inference under partial CSI. The piloted set is a strict subset of the
activated set, making the pilot budget a decision variable rather than a fixed
cost --- to our knowledge the first such treatment for FAS.

\item We give a concrete agent: a Gaussian belief over all $N$ ports from a
spatial prior and an order-$p$ autoregressive temporal model, with ports and
pilots chosen by greedy EFE minimization at $\mathcal{O}(NM)$ per slot, and an
exact reduced-rank belief that makes a $21\times21$ grid tractable.

\item We show that the agent reaches 91\% of a full-CSI genie's sum rate while
measuring $2.3\%$ of the ports, and degrades gracefully as pilots are further
withdrawn from ports it continues to serve.
\end{enumerate}

\section{Fluid Antenna System Model}
\label{sec:model}

\begin{figure}[t]
  \centering
%
%
%
\definecolor{cAct}{HTML}{1F6FB2}     
\definecolor{cActBg}{HTML}{E7F0F8}
\definecolor{cPil}{HTML}{D55E00}     
\definecolor{cPilBg}{HTML}{FCEFE4}
\definecolor{cCand}{HTML}{C3CAD3}    
\definecolor{cNeut}{HTML}{5E6A75}    
\definecolor{cNeutBg}{HTML}{F2F4F7}
\definecolor{cArrow}{HTML}{4A5560}

\begin{tikzpicture}[
  x=1cm, y=1cm,
  font=\scriptsize,
  >={Stealth[length=1.5mm]},
  stage/.style={
    draw=cNeut, fill=cNeutBg, rounded corners=1.2pt, align=center,
    inner sep=1.5pt, minimum height=6.6mm, minimum width=15.5mm,
    line width=0.45pt},
  actbox/.style={stage, draw=cAct, fill=cActBg, line width=0.6pt},
  pilbox/.style={stage, draw=cPil, fill=cPilBg, line width=0.6pt},
  sub/.style={font=\fontsize{5.6}{6.6}\selectfont, align=center, inner sep=1pt},
  key/.style={font=\fontsize{5.8}{6.6}\selectfont, inner sep=1pt, anchor=west},
]

\begin{scope}[shift={(0,2.55)}]
  \foreach \c in {0,...,6}
    \foreach \r in {0,...,6}
      \fill[cCand] (0.24*\c, 0.24*\r) circle (0.030);

  \foreach \p in {(0.24,1.20), (0.96,1.44), (0.48,0.48), (1.20,0.72), (0.72,0)}
    \fill[cAct] \p circle (0.062);

  \foreach \p in {(0.24,1.20), (1.20,0.72), (0.72,0)}
    \draw[cPil, line width=0.7pt] \p circle (0.125);

  \coordinate (unpiloted) at (0.48,0.48);
\end{scope}

\fill[cCand] (2.05,3.87) circle (0.030);
\node[key] at (2.20,3.87) {one of $N$ candidate ports};

\fill[cAct] (2.05,3.54) circle (0.062);
\node[key] at (2.20,3.54)
  {\textcolor{cAct}{\textbf{activated}}, $\mathcal{S}_t$, $|\mathcal{S}_t|=M$
   --- carries data};

\fill[cAct] (2.05,3.21) circle (0.062);
\draw[cPil, line width=0.7pt] (2.05,3.21) circle (0.125);
\node[key] at (2.20,3.21)
  {\textcolor{cPil}{\textbf{piloted}}, $\mathcal{Q}_t\subseteq\mathcal{S}_t$,
   $|\mathcal{Q}_t|=m\le M$};

\node[font=\fontsize{5.8}{7}\selectfont, anchor=north west,
      text width=6.15cm, align=left, inner sep=1pt] (callout) at (2.20,2.60)
  {the $M-m$ activated-but-unpiloted ports are \emph{served without ever being
   measured}, inferred through $\mathbf{R}$ and the AR($p$) prior};

\draw[->, line width=0.4pt, cPil]
  (callout.west) to[out=180, in=-20] ($(unpiloted)+(0.11,-0.02)$);

\node[stage]  (predict) at (0.875,1.10) {predict\\[-1pt]belief};
\node[actbox] (select)  at (2.575,1.10) {select $\mathcal{S}_t$};
\node[pilbox] (sense)   at (4.275,1.10) {pilot $\mathcal{Q}_t$};
\node[stage]  (update)  at (5.975,1.10) {update\\[-1pt]belief};
\node[stage]  (tx)      at (7.675,1.10) {precode\\[-1pt]\& transmit};

\draw[->, cArrow] (predict) -- (select);
\draw[->, cArrow] (select)  -- (sense);
\draw[->, cArrow] (sense)   -- (update);
\draw[->, cArrow] (update)  -- (tx);

\node[sub, below=1.0mm of predict] {AR($p$)\\[-1pt]dynamics};
\node[sub, below=1.0mm of select, text=cAct]
  {greedy EFE\\[-1pt]$\mathcal{O}(NM)$};
\node[sub, below=1.0mm of sense, text=cPil]
  {\emph{uplink}\\[-1pt]$m$ pilots};
\node[sub, below=1.0mm of update]  {Kalman\\[-1pt]all $N$ ports};
\node[sub, below=1.0mm of tx]      {\emph{downlink}\\[-1pt]$M$ ports};

\draw[->, dashed, line width=0.4pt, cNeut]
  (tx.north) -- (7.675,1.78)
  -- node[sub, above, pos=0.5, text=cNeut] {belief carried to slot $t+1$}
  (0.875,1.78) -- (predict.north);

\end{tikzpicture}
  \caption{Per-slot operation. The base station activates $M$ of $N$ candidate
  ports, but spends pilots on only $m\le M$ of them; the remaining $M-m$
  activated ports carry data while never being measured, their channels
  supplied by the belief alone. Both the activated set $\mathcal{S}_t$ and the
  piloted subset $\mathcal{Q}_t$ are chosen by minimizing the same expected
  free energy, which is what turns the pilot budget from a fixed cost into a
  design variable.}
  \label{fig:timeline}
\end{figure}
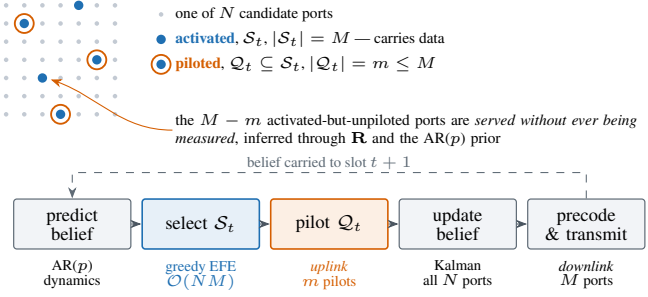

We consider the downlink of a single-cell multiuser system in which a base
station (BS) equipped with a fluid antenna serves $K$ single-antenna users.

\subsection{Array Geometry and Activation}
The BS fluid antenna comprises $N=N_x\times N_y$ candidate ports on a uniform
planar grid over a compact aperture with sub-half-wavelength spacing. It has
$M$ radio-frequency (RF) chains, so in slot $t$ it activates a subset
$\mathcal{S}_t\subseteq\{1,\dots,N\}$ with $|\mathcal{S}_t|=M$; only activated
ports radiate and connect to the front end. Serving $K$ streams requires
$N\ge M\ge K$. For an index set $\mathcal{A}$,
$\bP_{\mathcal{A}}\in\{0,1\}^{|\mathcal{A}|\times N}$ denotes the selection
matrix extracting the entries of a vector on $\mathcal{A}$.

\subsection{Spatio-Temporal Channel Model}
Let $\bh_k(t)\in\bbC^{N}$ be the channel from the $N$ ports to user $k$ in slot
$t$. The ports being densely spaced, their coefficients are spatially
correlated; under the classical Jakes model for rich scattering \cite{jakes1974microwave}
the correlation
depends only on the separation $d_{ij}$,
\begin{equation}
  R_{ij} = J_0\!\left(\frac{2\pi d_{ij}}{\lambda}\right),
  \label{eq:R}
\end{equation}
with $J_0(\cdot)$ the zeroth-order Bessel function of the first kind and
$\lambda$ the wavelength. Collecting these into $\bR$, the small-scale fading
is correlated Rayleigh, $\bh_k\sim\CN(\mathbf{0},\beta_k\bR)$, with $\beta_k$
the large-scale gain of user $k$. Sub-wavelength sampling makes $\bR$ strongly
rank-deficient, a fact Section~\ref{sec:aif} exploits.

The channel also ages between slots. Rather than the first-order Gauss--Markov
model usual in this literature, we retain $p$ past slots,
\begin{equation}
  \bh_k(t) = \sum_{i=1}^{p} a_i\,\bh_k(t-i) + \mathbf{e}_k(t),
  \quad
  \mathbf{e}_k(t)\sim\CN(\mathbf{0},\sigma_v^2\beta_k\bR),
  \label{eq:ar}
\end{equation}
with $\mathbf{e}_k(t)$ independent of the past. The coefficients
$\mathbf{a}=[a_1,\dots,a_p]^{\mathrm{T}}$ and innovation variance $\sigma_v^2$
solve the Yule--Walker system $\boldsymbol{\Gamma}\mathbf{a}=\mathbf{r}$,
$\sigma_v^2=r(0)-\mathbf{a}^{\mathrm{T}}\mathbf{r}$, for the Jakes temporal
autocorrelation $r(\tau)=J_0(2\pi f_D T_s\tau)$, where
$[\boldsymbol{\Gamma}]_{ij}=r(|i-j|)$, $f_D$ is the maximum Doppler frequency
and $T_s$ the slot duration; $p=1$ recovers the familiar first-order model.

\subsection{Pilot Allocation Under Partial CSI}
The BS cannot observe the full channel: a port yields information only when it
is measured. Prior port-selection work pilots every activated port, spending
$M$ pilot resources per slot. We separate the two decisions: in slot $t$ the BS
pilots a subset $\mathcal{Q}_t\subseteq\mathcal{S}_t$ with
$|\mathcal{Q}_t|=m\le M$ of the ports it has activated, observing
\begin{equation}
  \by_k(t) = \bP_{\mathcal{Q}_t}\bh_k(t) + \bn_k(t),
  \quad
  \bn_k(t)\sim\CN(\mathbf{0},\sigma_e^2\bI_m),
  \label{eq:obs}
\end{equation}
with $\sigma_e^2$ set by the pilot SNR. The unmeasured ports fall into two
classes. Those outside $\mathcal{S}_t$ must be inferred if the BS is ever to
select them. The $M-m$ ports in
$\mathcal{S}_t\setminus\mathcal{Q}_t$ are more unusual: they carry data in this
very slot while contributing no measurement, their channels supplied entirely
by the belief through \eqref{eq:R} and \eqref{eq:ar}
(Fig.~\ref{fig:timeline}). Setting $m<M$ turns the pilot budget from a fixed
overhead into a quantity the agent chooses.

\subsection{Hybrid Transmission and Achievable Rate}
The $M$ activated ports are driven by $n_{\mathrm{RF}}\le M$ RF chains through
a fully-connected network of unit-modulus phase shifters, giving the effective
precoder
\begin{equation}
  \bW = \bF_{\mathrm{RF}}\bW_{\mathrm{BB}}\in\bbC^{M\times K},
  \quad
  \big|[\bF_{\mathrm{RF}}]_{ij}\big|=1,
  \quad
  \|\bW\|_F^2\le P .
  \label{eq:hybrid}
\end{equation}
Writing $\bh_{k,\mathcal{S}}=\bP_{\mathcal{S}}\bh_k$ and treating inter-user
interference as noise, user $k$ achieves
\begin{equation}
  R_k = \log_2\!\left(
    1+\frac{\big|\bh_{k,\mathcal{S}}^{\mathrm{H}}\bw_k\big|^2}
           {\sum_{j\neq k}\big|\bh_{k,\mathcal{S}}^{\mathrm{H}}\bw_j\big|^2+\sigma^2}
  \right),
  \label{eq:rate}
\end{equation}
with $\sigma^2$ the receiver noise power. The target precoder is a regularized
MMSE filter built from the belief rather than from measured CSI, so it accounts
for residual uncertainty --- dominant at the $M-m$ unpiloted ports --- which
$\bF_{\mathrm{RF}}\bW_{\mathrm{BB}}$ then factorizes by alternating
minimization. For a fully-connected network with infinite-resolution shifters,
$n_{\mathrm{RF}}\ge2K$ chains represent any digital precoder exactly
\cite{sohrabi2016hybrid,zhang2005variable}.

\subsection{Switching Cost and Problem Formulation}
Reconfiguring between slots costs delay and energy \cite{fas_dropvelocity},
charged through the number
of ports that change, $C_t=|\mathcal{S}_t\,\triangle\,\mathcal{S}_{t-1}|$ with
$\triangle$ the symmetric set difference. The design problem is
\begin{equation}
  \begin{aligned}
    \max_{\{\mathcal{S}_t,\,\mathcal{Q}_t\}}\;\;
      & \sum_{t=1}^{T}\left(\sum_{k=1}^{K}R_k(t)-\eta_{\mathrm{sw}}C_t\right)\\[-2pt]
    \text{s.t.}\;\;
      & |\mathcal{S}_t|=M,\quad
        \mathcal{Q}_t\subseteq\mathcal{S}_t,\quad
        |\mathcal{Q}_t|=m,
  \end{aligned}
  \label{eq:problem}
\end{equation}
with $\eta_{\mathrm{sw}}\ge0$ trading throughput against reconfiguration cost.
Its difficulty is that $R_k(t)$ depends on the channel at ports that
\eqref{eq:obs} never reveals --- the problem we address next through active
inference.

\section{Active Inference for Port Selection and Pilot Allocation}
\label{sec:aif}

We now cast the problem as active inference. The agent holds a generative model
of the channel, maintains a Bayesian belief over all $N$ ports (perception),
and each slot chooses the activated set and the pilot subset that minimize an
expected free energy (action). The generative model needs no new ingredients:
the prior is $\CN(\mathbf{0},\beta_k\bR)$, the transition is the AR($p$)
dynamics \eqref{eq:ar}, and the likelihood is the partial observation
\eqref{eq:obs}. Preferences complete it, and here they are not a hand-tuned
vector as in textbook active inference but the communication utility itself ---
the rate net of switching in \eqref{eq:problem} --- so there is no arbitrary
preference to set.

\subsection{Perception: Complex Kalman Belief}
Stacking the last $p$ snapshots into
$\bz_k(t)=[\bh_k(t)^{\mathrm{T}},\dots,\bh_k(t-p+1)^{\mathrm{T}}]^{\mathrm{T}}
\in\bbC^{pN}$ turns \eqref{eq:ar} into a first-order recursion
$\bz_k(t)=\bF\bz_k(t-1)+\boldsymbol{\epsilon}_k(t)$, with
$\bF=\mathbf{C}_{\mathbf{a}}\otimes\bI_N$ for $\mathbf{C}_{\mathbf{a}}$ the
companion matrix of $\mathbf{a}$, and process noise
$(\mathbf{e}_1\mathbf{e}_1^{\mathrm{T}})\otimes(\sigma_v^2\beta_k\bR)$ confined
to the current-time block. Observation \eqref{eq:obs} reads
$\by_k(t)=\mathbf{G}\bz_k(t)+\bn_k(t)$ with
$\mathbf{G}=[\,\bP_{\mathcal{Q}_t}\;\;\mathbf{0}\,]$: pilots see the current
channel only.

The model is linear-Gaussian, so the exact posterior is Gaussian,
$q(\bz_k)=\CN(\hat{\bz}_k,\bP_k)$, maintained by a per-user complex Kalman
filter. The predict step propagates the belief through the AR dynamics,
encoding channel aging,
\begin{equation}
  \hat{\bz}_k \leftarrow \bF\hat{\bz}_k,
  \qquad
  \bP_k \leftarrow \bF\bP_k\bF^{\mathrm{H}} + \mathbf{Q}_k ,
  \label{eq:predict}
\end{equation}
and the pilots on $\mathcal{Q}_t$ correct it through the Kalman gain
$\mathbf{K}_k=\bP_k\mathbf{G}^{\mathrm{H}}
(\mathbf{G}\bP_k\mathbf{G}^{\mathrm{H}}+\sigma_e^2\bI_m)^{-1}$,
\begin{align}
  \hat{\bz}_k &\leftarrow \hat{\bz}_k
     + \mathbf{K}_k\big(\by_k-\mathbf{G}\hat{\bz}_k\big),
  \label{eq:mean}\\
  \bP_k &\leftarrow
     \big(\bI-\mathbf{K}_k\mathbf{G}\big)\bP_k
     \big(\bI-\mathbf{K}_k\mathbf{G}\big)^{\mathrm{H}}
     + \sigma_e^2\mathbf{K}_k\mathbf{K}_k^{\mathrm{H}} .
  \label{eq:joseph}
\end{align}
Only the $m\times m$ innovation covariance is inverted, regularized by
$\sigma_e^2\bI_m$, so the rank-deficient $\bR$ needs no artificial jitter, and
the Joseph form \eqref{eq:joseph} keeps $\bP_k$ Hermitian positive
semidefinite. The belief over the current channel is the leading block,
$\bmu_k=\boldsymbol{\Pi}\hat{\bz}_k$ and
$\bSig_k=\boldsymbol{\Pi}\bP_k\boldsymbol{\Pi}^{\mathrm{H}}$ with
$\boldsymbol{\Pi}=[\,\bI_N\;\;\mathbf{0}\,]$. Two couplings then do the work,
and together they are what permit $m<M$: spatially the off-diagonals of $\bR$
carry a measurement to a port's neighbours, so $m$ pilots inform far more than
$m$ ports, and temporally \eqref{eq:predict} carries information across slots,
so a port piloted earlier retains a usable estimate now.

\subsection{Exact Reduced-Rank Belief}
A $1764\times1764$ covariance per user per slot is infeasible, and
unnecessary. Truncating $\bR=\bB\boldsymbol{\Lambda}\bB^{\mathrm{H}}$ to the $r$
eigenvalues carrying all but $10^{-6}$ of the energy, and noting
$\bh_k(t)\in\mathrm{range}(\bR)$, the factorization
$\bh_k(t)=\bB\,\mathbf{c}_k(t)$, $\mathbf{c}_k(t)\in\bbC^{r}$, holds
\emph{exactly}: $\mathbf{c}_k$ is a sufficient statistic. The AR dynamics act
identically on every port, so $\bI_N$ becomes $\bI_r$, the state falls from $pN$
to $pr$ and the cost from $\mathcal{O}((pN)^3)$ to $\mathcal{O}((pr)^3)$. At
$N=441$ the numerical rank is $r=26$: $4.9\times10^{3}$ less arithmetic at no
modelling cost.

\subsection{Action: Expected Free Energy}
The expected free energy of a policy is conventionally the sum of two terms: a
pragmatic term, the expected log-preference over outcomes, and an epistemic
term rewarding informative actions. Reconfiguration cost does not need a third
term, because preferences are a modelling choice and ours are simply that the
agent prefers outcomes with high rate and few port changes,
\begin{equation}
  \ln\tilde{p}\big(o_t\,|\,\mathcal{S}\big) \;=\;
    \sum_{k}R_k(t) \;-\; \eta_{\mathrm{sw}}
    \big|\mathcal{S}\,\triangle\,\mathcal{S}_{t-1}\big| \;+\; \mathrm{const},
  \label{eq:pref}
\end{equation}
which is exactly the per-slot integrand of the system objective
\eqref{eq:problem}. The reconfiguration count is deterministic given the
action, so its expectation is itself and the pragmatic value remains in closed
form. Scoring a candidate pair $(\mathcal{S},\mathcal{Q})$ therefore needs only
the canonical two terms,
\begin{equation}
  G(\mathcal{S},\mathcal{Q}) =
   -\Big[\mathrm{Prag}(\mathcal{S})
     -\eta_{\mathrm{sw}}\big|\mathcal{S}\,\triangle\,\mathcal{S}_{t-1}\big|\Big]
   -\,\beta_w\,\mathrm{Epis}(\mathcal{Q}),
  \label{eq:efe}
\end{equation}
with $\beta_w>0$ an exploration precision. The bracket is the expected
log-preference \eqref{eq:pref} under the belief. Its first part is the sum rate
predicted under the robust MMSE precoder, penalized by the belief's own
residual uncertainty: with
$\bmu_{k,\mathcal{S}}=\bP_{\mathcal{S}}\bmu_k$,
$\bSig_{k,\mathcal{S}}=\bP_{\mathcal{S}}\bSig_k\bP_{\mathcal{S}}^{\mathrm{H}}$,
\begin{equation}
  \mathrm{Prag}(\mathcal{S}) = \sum_{k}
    \log_2\!\left(1+
    \frac{\big|\bmu_{k,\mathcal{S}}^{\mathrm{H}}\bw_k\big|^2}{\mathcal{I}_k}
    \right),
  \label{eq:prag}
\end{equation}
\begin{equation}
  \mathcal{I}_k =
    \sum_{j\neq k}\big|\bmu_{k,\mathcal{S}}^{\mathrm{H}}\bw_j\big|^2
    + \sum_{j}\bw_j^{\mathrm{H}}\bSig_{k,\mathcal{S}}\bw_j
    + \sigma^2 .
  \label{eq:leak}
\end{equation}
The middle term of \eqref{eq:leak} makes the agent conservative when uncertain:
a port with large posterior variance contributes interference the precoder
cannot null. It bites hardest at the $M-m$ ports served without being piloted,
so \eqref{eq:prag} already prices the cost of not measuring them. The epistemic
value is the mutual information the pilots on $\mathcal{Q}$ would yield,
\begin{equation}
  \mathrm{Epis}(\mathcal{Q}) = \sum_{k}
    \log_2\det\!\left(\bI_{m}
      + \sigma_e^{-2}\,
      \bP_{\mathcal{Q}}\bSig_k\bP_{\mathcal{Q}}^{\mathrm{H}}\right).
  \label{eq:epis}
\end{equation}
Being a log-determinant rather than a sum of marginal variances,
\eqref{eq:epis} prices redundancy: two candidates that are both uncertain
\emph{and} strongly correlated carry nearly the same information, so the second
earns little once the first is measured.

\begin{algorithm}[t]
\caption{Active inference for joint port selection and pilot allocation (slot $t$)}
\label{alg:aif}
\begin{algorithmic}[1]
\REQUIRE belief $\{\hat{\bz}_k,\bP_k\}_{k=1}^{K}$, previous set
         $\mathcal{S}_{t-1}$, budgets $M$, $m$, weights
         $\beta_w,\eta_{\mathrm{sw}}$
\ENSURE  activated set $\mathcal{S}_t$, pilot set $\mathcal{Q}_t$, precoder $\bW$
\STATE $\hat{\bz}_k\!\leftarrow\!\bF\hat{\bz}_k$,\;
       $\bP_k\!\leftarrow\!\bF\bP_k\bF^{\mathrm{H}}\!+\!\mathbf{Q}_k$\;$\forall k$
       \hfill $\triangleright$ predict, \eqref{eq:predict}
\STATE $\mathcal{S}\leftarrow\emptyset$
\FOR{$i=1$ \TO $M$}
  \STATE $p^{\star}\!\leftarrow\!\argmax_{p\notin\mathcal{S}}
         \big[G(\mathcal{S})-G(\mathcal{S}\cup\{p\})\big]$
         \hfill $\triangleright$ \eqref{eq:efe}
  \STATE $\mathcal{S}\leftarrow\mathcal{S}\cup\{p^{\star}\}$
\ENDFOR
\STATE $\mathcal{S}_t\leftarrow\mathcal{S}$;\;\; $\mathcal{Q}\leftarrow\emptyset$
\FOR{$j=1$ \TO $m$}
  \STATE $q^{\star}\!\leftarrow\!\argmax_{q\in\mathcal{S}_t\setminus\mathcal{Q}}
         \mathrm{Epis}\big(\mathcal{Q}\cup\{q\}\big)$
         \hfill $\triangleright$ \eqref{eq:epis}
  \STATE $\mathcal{Q}\leftarrow\mathcal{Q}\cup\{q^{\star}\}$
\ENDFOR
\STATE $\mathcal{Q}_t\leftarrow\mathcal{Q}$
\STATE transmit pilots on $\mathcal{Q}_t$, receive $\by_k$
\STATE update the belief from the $m$-dim.\ innovation
       \hfill $\triangleright$ \eqref{eq:mean}, \eqref{eq:joseph}
\STATE $\bW\leftarrow\bF_{\mathrm{RF}}\bW_{\mathrm{BB}}$ from the updated belief
       \hfill $\triangleright$ \eqref{eq:hybrid}
\STATE transmit on $\mathcal{S}_t$; incur
       $\eta_{\mathrm{sw}}|\mathcal{S}_t\triangle\mathcal{S}_{t-1}|$
\end{algorithmic}
\end{algorithm}

\subsection{Greedy Selection of Both Sets}
Exact minimization of \eqref{eq:efe} is combinatorial in both arguments, so we
build the sets greedily, each step adding the port whose inclusion most
decreases $G$,
\begin{equation}
  p^{\star} = \argmax_{p\notin\mathcal{S}}\;
    \Big[\,G(\mathcal{S}) - G\big(\mathcal{S}\cup\{p\}\big)\,\Big],
  \label{eq:greedy}
\end{equation}
until $|\mathcal{S}_t|=M$, then building $\mathcal{Q}_t$ the same way by
greedily maximizing \eqref{eq:epis} over $\mathcal{Q}\subseteq\mathcal{S}_t$
(Algorithm~\ref{alg:aif}). This is sequential greedy, not a ranking: being a
log-determinant, \eqref{eq:epis} makes a port's worth depend on which ports
$\mathcal{Q}$ already holds. One objective governs both sets --- eliminating
$\mathcal{Q}$ by inner minimization leaves
$\mathcal{S}_t=\argmin_{\mathcal{S}}\min_{\mathcal{Q}\subseteq\mathcal{S}}G$,
which choosing the two in sequence approximates at $\mathcal{O}(NM+Mm)$ per
slot. As $\mathrm{Epis}$ is monotone submodular and the switching term modular,
greedy carries the $(1-1/e)$ guarantee \cite{nemhauser1978analysis} on the
information-driven part; in
practice it is far tighter, and $\binom{M}{m}$ here is small enough to check
exhaustively --- greedy is exactly optimal in $95\%$ of slots and never below
$0.9996$ of it. Order matters: we predict, select, pilot, update, then precode
(Fig.~\ref{fig:timeline}), so piloted ports carry a posterior error of about
$\sigma_e^2$ rather than a one-step aging floor.

\section{Numerical Results}
\label{sec:results}

\begin{figure}[t]
  \centering
  \includegraphics[width=\columnwidth]{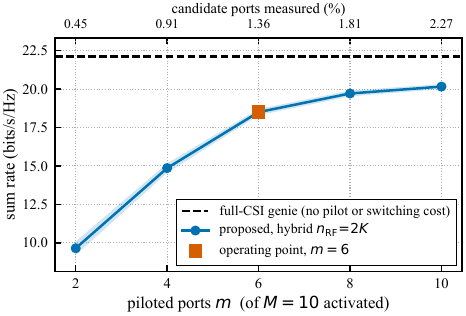}
  \caption{Sum rate against the pilot budget $m$ over 8 channel realizations;
  the band is the $95\%$ confidence interval. The upper axis restates $m$ as a
  fraction of the whole candidate grid. Withdrawing pilots from ports that
  continue to carry data costs little down to $m=6$, below which the belief can
  no longer cover the gap. The genie is a bound, not a scheme: it knows all $N$
  ports without spending pilots and moves between them for free.}
  \label{fig:pilots}
\end{figure}

\begin{table}[t]
\centering
\caption{Operating points; conditions as in Section~\ref{sec:results}-A, objective \eqref{eq:problem} at $\eta_{\mathrm{sw}}=1$. $m/M$ is the piloted fraction of the \textup{activated} ports, $m/N$ of the whole grid: even $m=M$ is partial CSI.}
\label{tab:ops}
\renewcommand{\arraystretch}{1.02}
\setlength{\tabcolsep}{3.2pt}
\footnotesize
\begin{tabular}{@{}ccccrrrr@{}}
\hline\hline
$m$ & $\eta_{\mathrm{sw}}$ & $m/M$ & $m/N$ & Sw./slot & Rate & \% genie & Obj. \\
 & & (\%) & (\%) & & (b/s/Hz) & & (b/s/Hz) \\
\hline
\multicolumn{8}{@{}l}{\emph{(a) pilot budget, at $\eta_{\mathrm{sw}}=1$}}\\
2 & $1$ & 20 & 0.45 & 1.01 & 9.65 & 43.6 & 8.64 \\
4 & $1$ & 40 & 0.91 & 3.76 & 14.86 & 67.2 & 11.10 \\
6\,$^\dagger$ & $1$ & 60 & 1.36 & 6.44 & 18.50 & 83.7 & 12.06 \\
8 & $1$ & 80 & 1.81 & 7.25 & 19.71 & 89.1 & 12.46 \\
10\,$^\ddagger$ & $1$ & 100 & 2.27 & 7.58 & 20.16 & 91.2 & 12.58 \\
\hline
\multicolumn{8}{@{}l}{\emph{(b) switching weight, at $m=6$}}\\
$6$ & $0$ & $60$ & $1.36$ & 18.89 & 17.61 & 79.6 & -1.27 \\
$6$ & $1$ & $60$ & $1.36$ & 6.44 & 18.50 & 83.7 & 12.06 \\
$6$ & $4$ & $60$ & $1.36$ & 0.00 & 18.01 & 81.4 & 18.01 \\
\hline
\multicolumn{4}{@{}l}{Full-CSI genie} & 15.34 & 22.12 & 100.0 & 6.78 \\
\hline\hline
\end{tabular}
\vspace{2pt}
\begin{flushleft}\footnotesize $^\dagger$ operating point used throughout; $^\ddagger$ every activated port piloted.\end{flushleft}
\end{table}

\subsection{Simulation Setup}
The BS serves $K=3$ single-antenna users from a $2\lambda\times2\lambda$ planar
fluid antenna with $N=441$ ports on a $21\times21$ grid, activating $M=10$ per
slot through $n_{\mathrm{RF}}=6=2K$ RF chains. Spatial correlation follows
\eqref{eq:R}; the channel evolves as a space-time Jakes process at $f_DT_s=0.1$
and the agent models it by \eqref{eq:ar} with $p=4$, fitted once by
Yule--Walker from the assumed Doppler spectrum. Transmit SNR is $15$\,dB,
$\rho=0.9$, $\beta_w=0.25$, and $T=40$ slots with all results averaged over the
second half, so the belief has settled. The \emph{full-CSI genie} used
throughout is a bound, not a competing scheme: it re-selects the best $M$ of
all $N$ ports greedily from perfect knowledge of the whole grid, spends no
pilots to acquire that knowledge, and pays no reconfiguration cost. No causal,
pilot-limited policy can attain it, and every percentage we quote is a fraction
of that ceiling. Finally, $m=M$ is the most any partial-sensing scheme can
measure without abandoning the budget.

\subsection{Pilot Budget}
Fig.~\ref{fig:pilots} and block~(a) of Table~\ref{tab:ops} report sum rate
against the pilot budget $m$. At $m=M=10$ the agent reaches
$20.16$\,bits/s/Hz, $91.2\%$ of the genie. The comparison deserves care: $m=M$
pilots every \emph{activated} port, but that set is $10$ of $441$, so only
$2.27\%$ of the array is measured and the other $97.7\%$ is inferred from $\bR$
and the temporal model. The gap to the genie is the price of never looking at
the rest of the grid, not of a coarse estimate.

Withdrawing pilots costs little over the useful range. At $m=6$ --- $40\%$
fewer pilot symbols --- the rate is $18.50$\,bits/s/Hz, $83.7\%$ of the genie,
so $7.5$ points of genie-relative rate buy back four of ten pilots. An
unpiloted port is not thereby unknown: it is predicted through \eqref{eq:ar}
and tightened by correlation with the piloted ports, since \eqref{eq:mean}
updates the whole reduced-rank state from an $m$-dimensional innovation. Below $m=6$ that mechanism runs out --- $14.86$ ($67.2\%$) at
$m=4$, $9.65$ ($43.6\%$) at $m=2$ --- a collapse rather than a decline: two
measurements cannot keep $r=26$ modes in step with the channel, so prediction
error outruns the AR model and the precoder inverts a covariance it no longer
trusts.

Block~(a) also shows the budget governing how much the array \emph{moves}:
switching rises monotonically with $m$, from $1.01$ ports per slot at $m=2$ to
$7.58$ at $m=M$. An agent that cannot see is also unwilling to act, because
\eqref{eq:efe} cannot certify that an unmeasured port is worth the
reconfiguration. The objective therefore rises more gently than the rate ---
$12.06$ at $m=6$ against $12.58$ at $m=M$ --- so $m=6$ retains $95.9\%$ of the
best attainable objective for $40\%$ fewer pilots, and we adopt it hereafter.

\subsection{The Switching Cost}
Block~(b) of Table~\ref{tab:ops} sweeps the switching weight
$\eta_{\mathrm{sw}}$ at $m=6$, and shows how little rate the objective
\eqref{eq:problem} trades away. Raising $\eta_{\mathrm{sw}}$ from $0$ to $4$ takes the agent from $18.89$
reconfigured ports per slot --- rebuilding the aperture every slot --- to none
at all, while the sum rate moves by $0.89$\,bits/s/Hz: an array that never
moves is within $2.7\%$ of one that moves constantly.

Rate is also not monotone in $\eta_{\mathrm{sw}}$. It \emph{peaks} at
$\eta_{\mathrm{sw}}=1$ ($18.50$\,bits/s/Hz) rather than at
$\eta_{\mathrm{sw}}=0$ ($17.61$): charging for movement makes the array faster.
Under partial CSI a retained port accumulates posterior precision across slots,
whereas an abandoned one takes its belief with it and its replacement starts
from the prediction alone, so an unpenalized agent chases marginally better
ports and pays in estimation quality. Some reluctance to move is an information
policy, not merely a concession to the hardware --- a partial-CSI effect
invisible to any formulation that assumes the channel is known before the port
is chosen.

Finally, the genie loses on the objective. It leads by $3.64$\,bits/s/Hz in raw
rate but reconfigures $15.34$ ports per slot, so at $\eta_{\mathrm{sw}}=1$ its
objective is $6.78$ against our $18.01$. Equating the two,
$(22.12-18.01)/(15.34-0.00)$, the genie is beaten for any
$\eta_{\mathrm{sw}}>0.27$: unless reconfiguring a port is worth less than a
quarter of a bit per second per hertz, perfect CSI with no movement discipline
is the worse policy. This is the sense in which the switching term earns its
place inside the expected free energy rather than being appended to it.

\begin{figure}[t]
  \centering
  \includegraphics[width=\columnwidth]{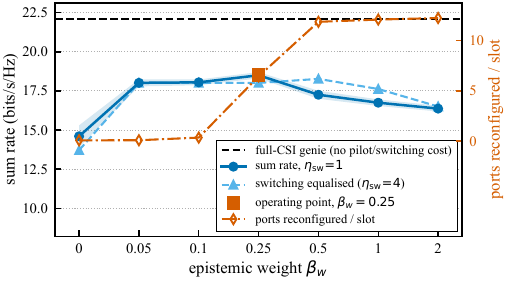}
  \caption{Epistemic ablation at $m=6$ over 8 realizations: everything is held
  fixed and only the epistemic weight $\beta_w$ is varied, so $\beta_w=0$ is
  pure rate-greedy selection with no active inference; it is stranded at
  $66.0\%$ of the genie, against $83.7\%$ at the marked operating point. The
  genie again pays neither pilots nor switching. The dashed curve repeats the
  sweep at $\eta_{\mathrm{sw}}=4$, where every policy is driven to zero
  reconfiguration (Table~\ref{tab:ops}), so the two arms move identically and
  the remaining gap cannot be attributed to movement.}
  \label{fig:ablation}
\end{figure}
\subsection{Does the Epistemic Term Earn Its Place?}
Active inference would be an empty relabelling of rate-greedy selection if the
epistemic term did no work, so we ablate it: Fig.~\ref{fig:ablation} holds
belief, pilot rule, precoder and channel fixed and varies only $\beta_w$. At
$\beta_w=0$ the agent selects purely on predicted rate and is stranded at
$14.60$\,bits/s/Hz, $66.0\%$ of the genie; at $\beta_w=0.25$ it reaches
$18.50$, $83.7\%$.

A sceptic can reply that the exploring agent merely moves more, and that
movement buys the rate. The dashed curve settles it: at $\eta_{\mathrm{sw}}=4$
both agents are driven to \emph{zero} reconfiguration, so movement is identical
by construction, and the epistemic agent still leads by $4.30$\,bits/s/Hz ---
$19.4$ points of genie-relative rate --- while mean posterior variance across
all $N$ ports falls from $0.72$ to $0.37$. The term buys knowing \emph{where to
look}, not looking more.

The optimum is interior: beyond $\beta_w\approx0.25$ switching saturates near
$12$ ports per slot and rate falls to $74\%$ of the genie by $\beta_w=2$. We
take $\beta_w=0.25$, which maximizes rate. A smaller $\beta_w=0.05$ does score
higher on the $\eta_{\mathrm{sw}}=1$ objective ($17.94$ against $12.06$), but
only by reconfiguring $0.08$ ports per slot --- a static array, not a fluid
one. The objective rewards standing still; the point of a fluid antenna is that
it need not.

\subsection{Hybrid Transmission}
All results use $n_{\mathrm{RF}}=6=2K$ RF chains, not $M=10$. At this
threshold the hybrid precoder \eqref{eq:hybrid} is digital-exact: against a
fully digital front end the loss is below $0.001$\,bits/s/Hz at every pilot
budget in Table~\ref{tab:ops}. The $40\%$ saving in RF chains is free, so the
results above are not artefacts of a relaxed transmit architecture.

\section{Conclusion}
\label{sec:conc}
We cast joint port selection and pilot allocation for a fluid antenna as active
inference, deriving a two-term expected free energy from an explicit preference
density in which reconfiguration cost is a preference over outcomes rather than
a third, ad hoc penalty. The agent attains $91.2\%$ of a full-CSI genie's sum
rate while measuring $2.3\%$ of the candidate ports and $83.7\%$ while
measuring $1.4\%$, and beats that genie on the reconfiguration-aware objective
for any switching price above $0.27$\,bits/s/Hz per port. Ablating the
epistemic term costs $19.4$ points of genie-relative rate at equal switching,
so the information term is load-bearing rather than nominal. Both sets are
chosen greedily against a monotone submodular epistemic value at
$\mathcal{O}(NM+Mm)$ per slot.

Two extensions are out of scope here: the AR coefficients are fitted once from
an assumed Doppler spectrum, and learning them online so the agent tracks an
unknown or time-varying $f_D$ is left to future work, as is learning the
spatial correlation $\bR$ rather than assuming it from the array geometry.

\bibliographystyle{IEEEtran}
\bibliography{refs}

\end{document}